\documentclass[9pt, conference]{IEEEtran}

\ifCLASSINFOpdf
\else
\fi
\usepackage{multicol}
\usepackage{multirow}
\usepackage{cite}
\usepackage{amsmath,amssymb,amsfonts}
\usepackage{algorithmic}
\usepackage{graphicx}
\usepackage{textcomp}
\usepackage{xcolor}
\usepackage{subcaption}
\usepackage[lined,boxed,commentsnumbered,ruled]{algorithm2e}
\usepackage{makecell}

\usepackage{comment}
\usepackage{hyperref}

\begin{document}
%
\title{FaRAccel: FPGA-Accelerated Defense Architecture for Efficient Bit-Flip Attack Resilience in Transformer Models\vspace{-6ex}}



\author{\IEEEauthorblockN{\\ \\ \\ \\Najmeh Nazari,
Banafsheh Saber Latibari,
Elahe Hosseini, \\
Fatemeh Movafagh,
Chongzhou Fang,
Hosein Mohammadi Makrani,
Kevin Immanuel Gubbi,\\
Abhijit Mahalanobis,
Setareh Rafatirad,
Hossein Sayadi,
Houman Homayoun}
\IEEEauthorblockA{
Department of Electrical and Computer Engineering, University of California, Davis, CA, USA\\
Department of Electrical and Computer Engineering, University of Arizona, Tucson, AZ, USA \\
Department of Computer Engineering and Computer Science, California State University, Long Beach, Long Beach, CA, USA \\
Department of Computer Science, Simon Fraser University, Vancouver, Canada\\
Emails: \{nnazari, ehosseini, czfang, hmakrani, kgubbi, srafatirad, hhomayoun\}@ucdavis.edu}, \{banafsheh, amahalan\}@arizona.edu , hossein.sayadi@csulb.edu, fma44@sfu.ca
}



\maketitle

\begin{abstract}
Forget and Rewire (FaR) methodology has demonstrated strong resilience against Bit-Flip Attacks (BFAs) on Transformer-based models by obfuscating critical parameters through dynamic rewiring of linear layers. However, the application of FaR introduces non-negligible performance and memory overheads, primarily due to the runtime modification of activation pathways and the lack of hardware-level optimization. To overcome these limitations, we propose FaRAccel, a novel hardware accelerator architecture implemented on FPGA, specifically designed to offload and optimize FaR operations. FaRAccel integrates reconfigurable logic for dynamic activation rerouting, and lightweight storage of rewiring configurations, enabling low-latency inference with minimal energy overhead. We evaluate FaRAccel across a suite of Transformer models and demonstrate substantial reductions in FaR inference latency and improvement in energy efficiency, while maintaining the robustness gains of the original FaR methodology. To the best of our knowledge, this is the first hardware-accelerated defense against BFAs in Transformers, effectively bridging the gap between algorithmic resilience and efficient deployment on real-world AI platforms. 

\end{abstract}

\section{Introduction}
\label{sec:introduction}

The widespread deployment of Transformer-based architectures in natural language processing, security \cite{saber2025transformers, latibari2024transformers, ghimire2025hwrex, latibari2024automated, movafaghcyber}, health \cite{hosseini2025large}, computer vision \cite{latibari2025optimizing, saber2024iret}, and multimodal tasks \cite{fang2024large, shahir2025gist} has brought unprecedented advancements in machine learning. However, these models are increasingly being exposed to sophisticated adversarial threats targeting their parameters at the hardware level \cite{nazari2024llm}. Bit-Flip Attacks (BFAs), in particular, pose a severe risk by exploiting memory vulnerabilities to alter critical bits in model weights, leading to dramatic drops in accuracy with minimal perturbations \cite{nazari2024forget}. While various algorithmic defenses have been proposed, they often overlook the computational and latency constraints associated with real-world deployment, especially in edge-AI platforms.

To address the growing concern of BFA resilience in Transformer-based models, the Forget and Rewire (FaR) methodology was recently introduced \cite{nazari2024forget}. FaR mitigates BFAs by dynamically redistributing critical parameter responsibilities across less sensitive neurons, thereby obfuscating parameter importance and weakening the effectiveness of gradient-based attack algorithms. Experimental results have shown that FaR can reduce attacker success by up to 4× with negligible accuracy loss. However, FaR’s runtime overhead, induced by dynamic rewiring and activation rebalancing, remains a key challenge, particularly for time-sensitive applications on resource-constrained devices.

Current implementations of FaR rely on Python-based environments and software-level modifications to standard linear layers. This setup lacks optimization for low-latency inference and incurs non-trivial memory and compute overheads during runtime. More importantly, it fails to fully exploit the benefits of hardware reconfigurability and parallelism—critical attributes for deploying secure models in edge and embedded systems. Therefore, bridging the gap between the algorithmic robustness of FaR and the deployment efficiency required in practical settings is essential.

In this work, we propose FaRAccel, the first hardware accelerator specifically designed to implement the Forget and Rewire methodology. FaRAccel is built on an FPGA-based architecture and introduces a reconfigurable datapath for dynamically managing neuron connections at inference time. It integrates lightweight memory blocks for storing rewiring configurations and supports runtime sensitivity-aware routing, all without altering the model’s topology or retraining the network. This enables seamless integration with standard Transformer models.

We evaluate FaRAccel on a suite of Transformer architectures spanning both language and vision domains. Our results demonstrate that FaRAccel achieves substantial improvements in inference latency, up to 15× speedup compared to software-based FaR implementations, while preserving the same robustness against BFAs. These findings highlight the potential of architectural specialization in supporting resilient AI workloads.

To the best of our knowledge, FaRAccel is the first hardware-based defense mechanism specifically tailored for BFA mitigation in Transformers. Unlike previous works that rely on software-based obfuscation or defensive retraining, FaRAccel introduces a principled hardware-software co-design that ensures both robustness and efficiency. It thus opens a new frontier for integrating security-aware AI accelerators into edge and embedded ecosystems.

In summary, this paper makes the following key contributions: 
\begin{enumerate}
    \item  We propose FaRAccel, a novel FPGA-based accelerator for executing the Forget and Rewire operations in hardware;
    \item We design and implement a full-fledged hardware pipeline in FaRAccel, supporting dynamic rerouting, activation modulation, and parameter concealment;
    \item We conduct a comprehensive analysis and demonstrate substantial reductions in inference latency and energy consumption, while preserving the robustness against BFAs. 

    \item We present the first step toward secure and efficient deployment of Transformer models with algorithm-hardware co-design principles.
\end{enumerate}

The remainder of the paper is organized as follows. Section II reviews background on the topic. Section III motivates the need for an accelerator for FaR. Section IV details our methodology. Sections V and VI describe the FaRAccel architecture and implementation, and present the experimental setup and evaluation. Section VII surveys related work, and Section VIII concludes.

\section{Background}

In this section, we present a concise overview of the essential preliminary knowledge, especially on Transformers and bit flip attacks.

\subsection{Transformer-based Models}

The Transformer architecture underpins both Large Language Models and Vision Transformers, and is central to modern pipelines for natural language processing and image recognition \cite{hf}. In this section, we briefly review the Transformer to highlight why our method is compatible with—and effective for—this class of models.

\begin{figure}[!t]
\centering
\includegraphics[width=0.45 \textwidth] {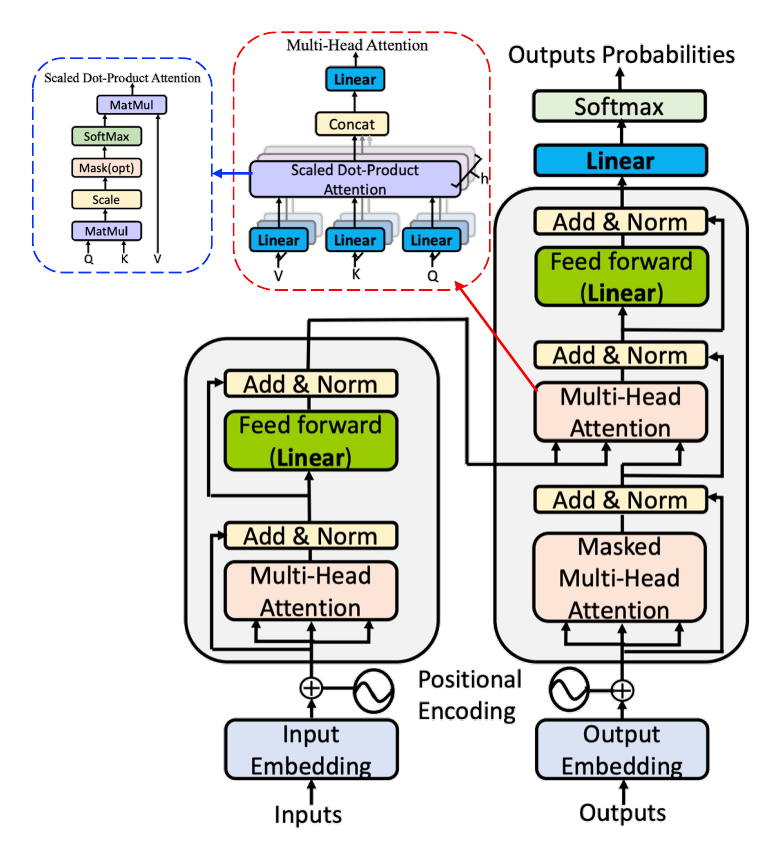}
\caption{Architecture of Transformers.}

\label{fig:trans}
\end{figure}

The canonical Transformer comprises two components \cite{vaswani2017attention}: an \emph{encoder} and a \emph{decoder} (Figure \ref{fig:trans}). The encoder maps input tokens to context-rich representations (features), enabling the model to “understand” the input. The decoder consumes these representations—together with appropriate conditioning—to produce an output sequence. Depending on the task, either component can be used on its own:

\begin{itemize}
\item \textbf{Encoder-only:} Best for understanding tasks (e.g., text classification, named entity recognition).
\item \textbf{Decoder-only:} Best for open-ended generation.
\item \textbf{Encoder–decoder:} Best for conditional generation (e.g., summarization, translation).
\end{itemize}

A defining feature of Transformers is the \emph{attention} mechanism, which selectively weights interactions among tokens. Attention encourages the model to emphasize informative tokens while downweighting less relevant ones when constructing each token’s representation. Attention masks in the encoder/decoder can further restrict which positions are visible (e.g., to ignore special or padded tokens, or to enforce causality).

In multi-head attention, tokens are first projected via three learned Linear mappings into \emph{Query} (Q), \emph{Key} (K), and \emph{Value} (V) vectors \cite{vaswani2017attention}. Given these projections, the attention operation computes similarity between Q and K to produce weights that are applied to V, yielding updated token representations that incorporate contextual information.

These computations are performed in parallel across multiple \emph{heads}. Concretely, the Q, K, and V projections are split into $H$ parts and processed independently, after which the head outputs are concatenated and linearly transformed to form the final attention output. This \emph{multi-head} design allows the model to capture diverse relationships and nuances across tokens, contributing to the strong performance of Transformers and aligning well with our approach.

\textbf{Why Transformers scale well.} Beyond attention, the building blocks of each layer are predominantly \emph{linear} operations: the Q/K/V projections and output projection in attention, and the position-wise feed-forward network (typically two Linear layers with a simple activation such as GELU). These are composed with residual connections and layer normalization, both of which are lightweight and highly parallelizable. Because training and inference are dominated by dense matrix multiplies (GEMMs), Transformers map efficiently to modern accelerators and libraries (e.g., cuBLAS/oneDNN), making them straightforward to scale by widening hidden dimensions, increasing head counts, and stacking more layers. The simplicity and uniformity of these linear components improve hardware utilization, enable predictable memory/computation profiles, and contribute to stable optimization at large batch sizes and model scales—factors that align well with our approach.

\subsection{Bit-Flip Attacks}


System robustness and safety are fundamentally dependent on strong memory isolation enforced by both software and hardware \cite{frassetto2018imix}. Yet even state-of-the-art DRAM can violate this isolation due to read-disturbance effects. An important example is RowHammer \cite{mutlu2019rowhammer}, where repeatedly activating and precharging (“hammering”) a DRAM row causes bit flips in physically adjacent victim rows. More recently, RowPress \cite{luo2023rowpress} demonstrated that, even on DDR4 systems hardened with RowHammer mitigations, user-level code can still breach isolation by simply holding a row open for an extended period; this sustained activation perturbs neighboring rows enough to induce bit flips.

\begin{figure}[!t]
\centering
\includegraphics[width=0.5 \textwidth] {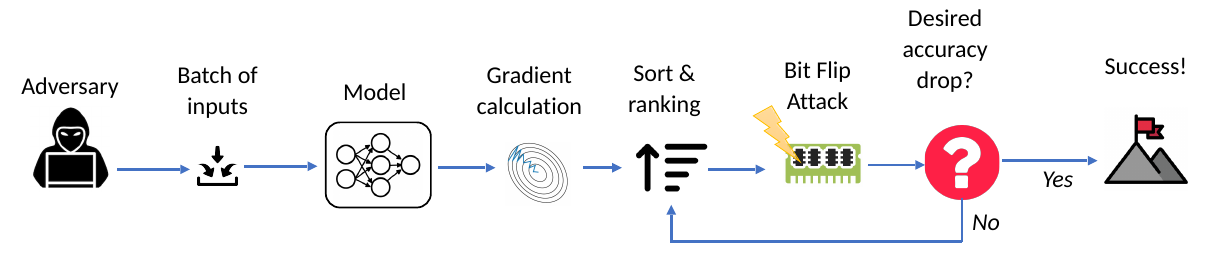}
\caption{ Bit-Flip Attack overview.}

\label{fig:bfa}
\end{figure}

Bit-flip attacks seek to exploit practical memory fault-injection mechanisms to degrade a model’s performance while changing as few bits as possible. Formally, the adversary operates under a strict flip budget and aims to identify a sparse set of bit locations whose alteration maximizes misclassification impact. In practice, this often means prioritizing parameters whose perturbation most strongly shifts decision boundaries, so that each flip yields an outsized effect.
A stealthy adversary further constrains this objective: the goal is to redirect inputs from a specific source class $p$ to a chosen target class $q$ ($q \neq p$) while leaving predictions for all other inputs essentially unchanged. By preserving overall accuracy—and thus the model’s external behavior—the attack remains difficult to detect, even as it reliably induces targeted errors on the $p \rightarrow q$ subset.


One of the most effective BFA variants is DeepHammer \cite{yao2020deephammer}, which improves bit-selection by employing gradient-based Progressive Bit Search (PBS). As illustrated in Figure\ref{fig:bfa}, at the $k^{\text{th}}$ iteration DeepHammer first computes a gradient-based saliency over the model’s parameters and selects the top $n$ candidate bits. It then evaluates each candidate by virtually flipping that bit and measuring the resulting loss, forming a loss set $L$. Repeating this procedure layer by layer yields $n \times l$ candidates and their corresponding losses, from which DeepHammer commits the single flip that maximizes the loss. This process iterates until the attack objective is met or a flip budget is exhausted. While the original DeepHammer restricts itself to one committed flip per DRAM page, the threat model assumes an adversary capable of flipping multiple bits within a page, which strictly generalizes DeepHammer and represents a stronger attacker.

\section{Motivation}

Forget-and-Rewire (FaR) was introduced to meet the need of a practical defense to raise the number of required flips (i.e., the attack cost) while preserving model utility without retraining. Inspired by neuroplasticity, FaR redistributes the influence of important parameters by coupling them with less important (often near-inactive or "dead") neurons in the same linear layer. By "hiding" critical parameters behind newly rewired activation paths, FaR dilutes gradient salience, frustrating the attacker's search and forcing more bit flips to achieve the same damage. Conceptually, this rewiring increases the attacker's workload while keeping the layer's functional mapping nearly unchanged.  

Mechanistically, FaR identifies (i) the most sensitive parameter in a layer and (ii) one or more dead neurons. It then applies a Forget step (disconnect the dead path) and a Rewire step (split and reroute the activation feeding the critical weight toward the dead neuron), moderated by a division factor that balances robustness and accuracy. This process is repeated iteratively across layers to meet a desired robustness budget. Importantly, FaR is a one-shot, post-training modification; its small configuration can be stored securely on-chip and applied at deployment via a custom linear layer that adjusts connections on the fly.  

Empirically, FaR reduces BFA success by roughly 1.4 to 4.2$\times$ with minimal accuracy cost (less than 2\% across common benchmarks), and it composes naturally with detection/recovery schemes (e.g., NeuroPots), providing complementary benefits: FaR conceals and redistributes sensitivity; detectors watch for tampering. These characteristics make FaR attractive as a lightweight hardening step for off-the-shelf Transformer models.  

\begin{figure}[!t]
\centering
\includegraphics[width=0.48 \textwidth] {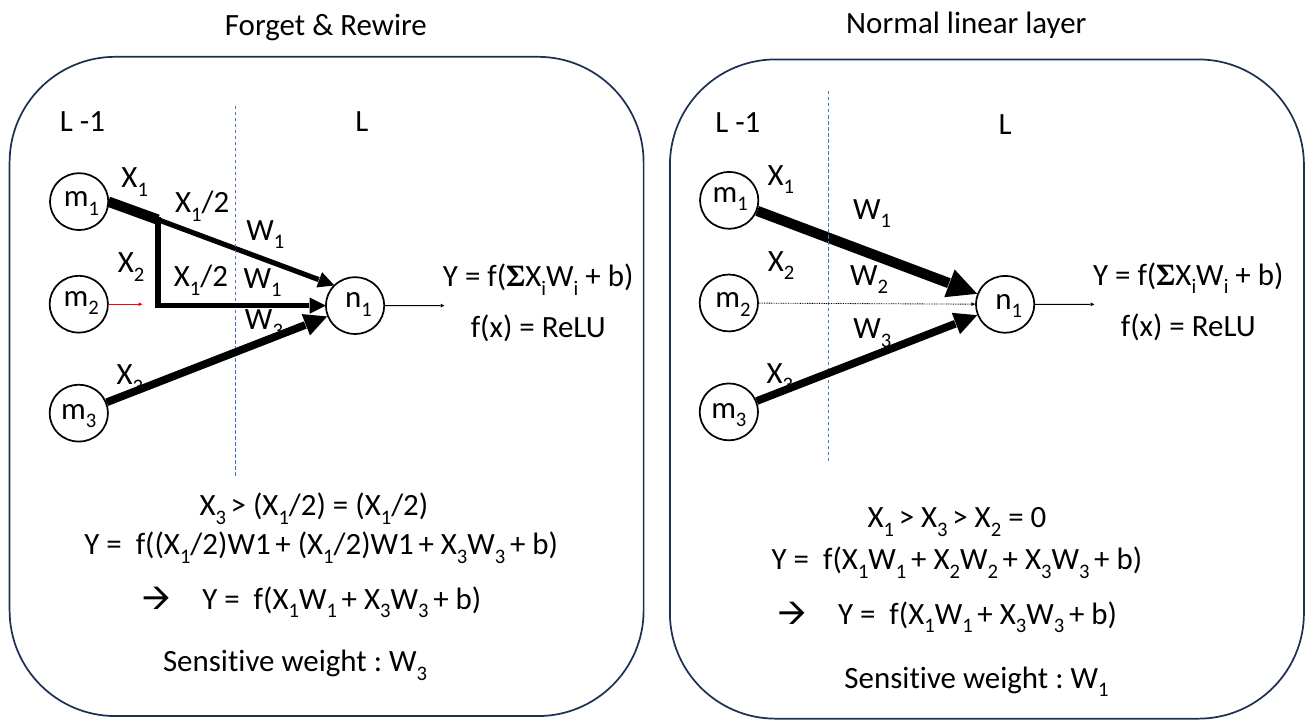}
\caption{ Forget and rewire example on a simple linear model \cite{nazari2024forget}}

\label{fig:far}
\vspace{-1em}
\end{figure}

Since the understanding of FaR's low level implementation is important to get insight into why its current impelemntation is not efficient and requires a solution, we explain its core functionality in the following.  
Figure \ref{fig:far} illustrates our \textit{Forget and Rewire} scheme. In Figure \ref{fig:far}(right side), we have a standard linear model where three neurons in layer $l-1$ are connected to one neuron in layer $l.$ Suppose we perform sensitivity analysis on three parameters of $W$ using a batch of input data. In this analysis, we find that $X_2$, the activation from neuron $m_2$, is predominantly close to 0, making $W_2$ the least important. Similarly, because $X_1$ is larger than $X_3$, the gradient of the output with respect to $W_1$ surpasses that of $W_3$. Furthermore, the output of neuron $Y$ depends solely on $X_1W_1 + X_3W_3$.

In Figure \ref{fig:far}(left side), we showcase the layer following the application of our Forget and Rewire scheme. Since $m_2$ is considered a dead neuron, its connection to $n_1$ becomes a candidate for the "Forget" operation. Recognizing that $W_1$ holds the utmost importance in the network, we opt to hide it from potential attackers. Conversely, we select $W_2$ for the "Rewiring" operation and replace its value with that of $W_1$ as $W_2$ no longer affects the output of neuron $n_1$. To achieve this, we divide the activation from neuron $m_1$ by 2 (the division factor). Half of this division is passed to the same parameter $W_1$, and the other half replaces the activation of the dead neuron, going to $W_2$ (which now holds the value of $W_1$). This way, when attackers employ gradient search and loss backpropagation, they discover that the weight ranking has changed. With $X_1$ now reduced by half, it is smaller than $X_3$, making $W_3$ the most important parameter in the layer. It's worth noting that despite modifying a weight's value and rewiring activations, the output of neuron $n_1$ remains consistent, and the layer's functionality undergoes minimal change.

If an attacker arbitrarily targets a critical parameter (known to the developer, such as $W_1$), the attack is not as effective as in the non-FaR model. This is because $W_2$, to which the importance of $W_1$ has been redistributed, also needs to be targeted.
Consequently, the model's resilience against random attacks is enhanced as well.

That said, FaR introduces measurable performance costs at inference time. Across evaluated models/datasets, FaR increases the inference time linearly when the number of parameters to rewire increases, and its time complexity is O(n). Crucially, the dominant runtime overhead stems from FaR’s custom linear layer implemented in Python and explained above, which lacks highly optimized binary kernels, unlike standard PyTorch linear ops. hence, the overhead grows with the number of FaR-enabled parameters and chosen division factor.   

\begin{figure}[!t]
\centering
\includegraphics[width=0.45\textwidth] {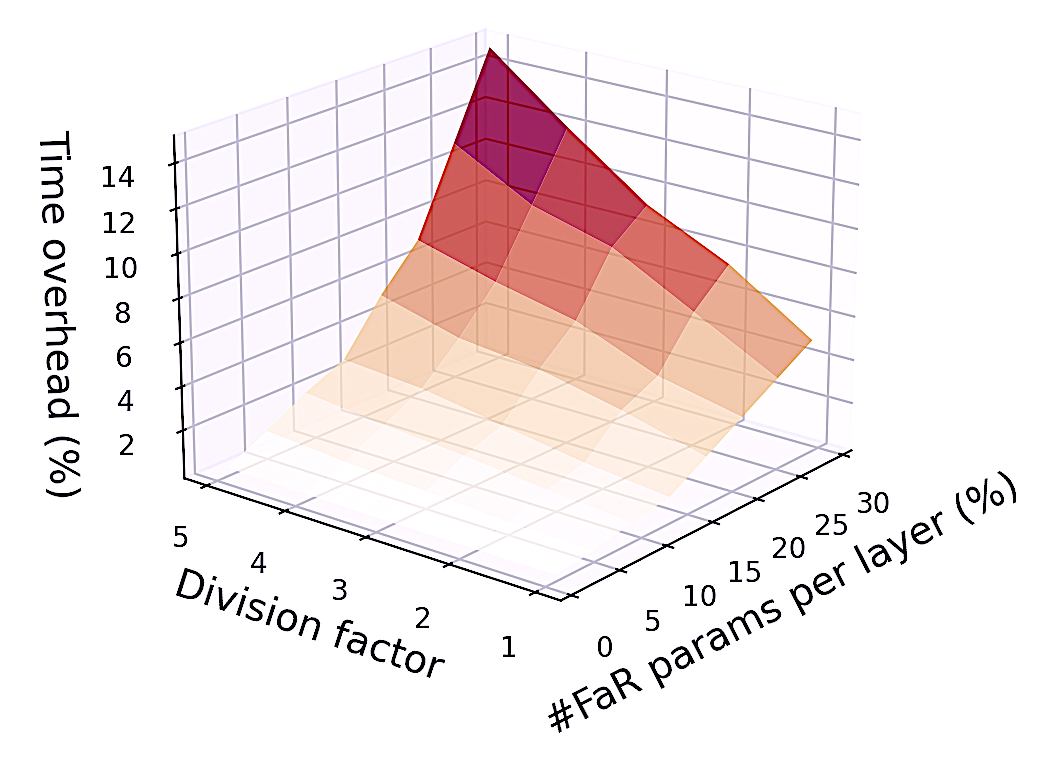}
 \vspace{-1em}
\caption{The impact of the number of FaR parameters and division factor per layer on the inference time overhead \cite{nazari2024forget}}

\label{fig:tim_far}

\end{figure} 

These software-centric bottlenecks motivate hardware support. If the very place where FaR "pays" its overhead is the linear layer (extra activation routing, scaling, and weight aliasing), then moving those micro-ops into a reconfigurable datapath can recover most of the cost. A hardware FaR unit can (i) fetch a compact, secure FaR configuration from on-chip memory, (ii) apply division-factor scaling and activation duplication in-flight, (iii) steer activations through a small crossbar/mux fabric before MACs, and (iv) write back results without extra framework-level bookkeeping. In other words, we fuse FaR into the linear layer's dataflow so rewiring is "free" in the steady state. This direction also aligns with the paper's own observation that an optimized, lower-level implementation is necessary to keep overhead acceptable as models scale. 

Taken together, these points motivate the FPGA-based FaRAccel architecture introduced next: a minimal, reconfigurable augmentation of the linear layer that provides low-latency activation rerouting, in-line scaling, and secure configuration, eliminating the software tax without changing the model's topology.

\section{Methodology and System Overview}

\begin{figure}[!t]
\centering
\includegraphics[width=0.5 \textwidth] {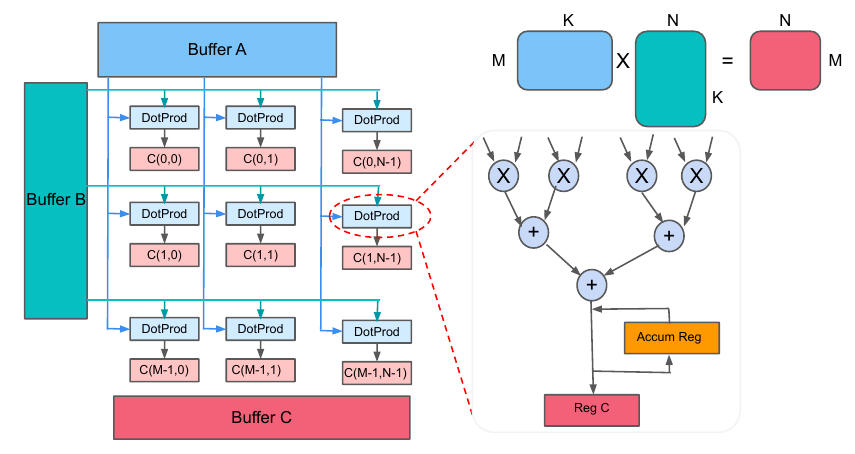}
\caption{Matrix multiplication that GeMM accelerator processes in one cycle with 3D spatial unrollings }
\label{fig:gemm}
\end{figure}

FaRAccel targets the cost of executing Forget-and-Rewire (FaR) during inference by turning FaR from a software-level graph rewrite into a lightweight operand-selection problem inside the accelerator’s matrix-multiply datapath. The design for GeMM block is presented in Figure \ref{fig:gemm}. The central idea is to preserve the baseline GEMM throughput when no rewiring applies, and to keep that same throughput when rewiring is enabled by moving all FaR decisions into a tiny configuration memory and a per-lane redirect network that sits in front of the multipliers. In this formulation, FaR no longer introduces extra arithmetic on the critical path. Instead, it changes which weight each lane consumes and whether that weight is a baseline value, a pre-scaled donor value, or zero. The methodology separates FaR into two main steps: an offline compilation step and a runtime application step. 

\subsection{Offline Step} A sensitivity analysis produces a compact configuration for each linear layer (or for each tile of a large layer). This configuration, which we call a \textit{FaRMap}, contains only exceptions relative to the baseline weight layout. For each output neuron, the map lists a small number of victim indices, bounded to 15\% of its inputs in our design, and for each victim, specifies whether the multiply is skipped or rewired to a donor index. If the rewire requires dividing the contribution, the division factor is restricted to two or three to simplify hardware. To avoid inserting dividers in the datapath, donor weights that require scaling are pre-materialized offline into a small “shadow” store as FP16 copies of the donor value multiplied by 1/2 or 1/3. The offline step therefore emits two tiny artifacts per layer: the FaRMap and the shadow weights.

\subsection{Runtime Step} 
\begin{figure}[!t]
\centering
\includegraphics[width=0.4 \textwidth] {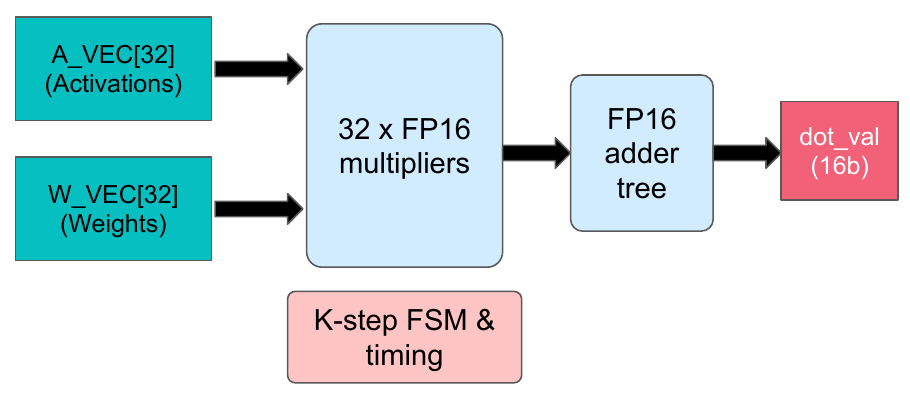}
\caption{Baseline DPE }
\label{fig:dpe}
\end{figure}

\begin{figure}[!t]
\centering
\includegraphics[width=0.5 \textwidth] {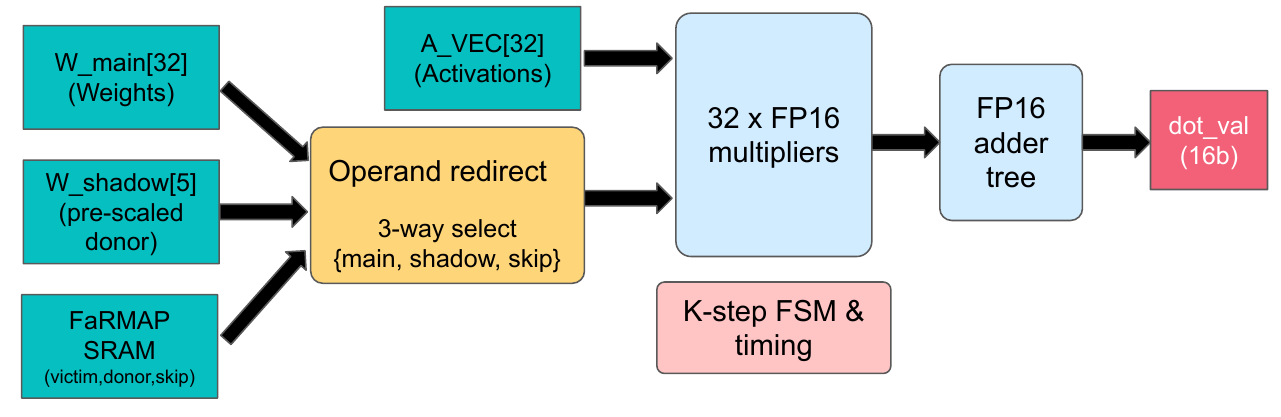}
\caption{FaR-aware DPE }
\label{fig:far_dpe}
\end{figure}

Operates on 32×32 FP16 tiles using an output-stationary dataflow. Activations and baseline weights are streamed into on-chip tile buffers, while the FaRMap and any referenced shadow weights are prefetched into small SRAMs. When an output row (or column) is issued, the corresponding slice of the FaRMap is also fed to the the controller. FaRMap synthesizes a 32-entry select vector that encodes, for each lane, whether to use the baseline weight, a shadow donor, or zero, and latches that vector for the duration of the row. The dot-product engines then consume activations and the effective weights at the same cadence as a vanilla GEMM. Because decision making is confined to a single cycle of select generation per row and the shadow and baseline weights are available on distinct read ports, the multiplier array is never stalled, and steady-state throughput is preserved \cite{guo2025transitive}. The diagram presented in Figure \ref{fig:dpe} is the baseline DPE engine, and Figure \ref{fig:far_dpe} illustrates the Far-aware DPE block designed to enable our proposed run-time operation.

Methodologically, this design preserves the trained model topology and weights. FaR can be enabled or disabled per layer at runtime without retraining, and deployment consists of loading the FaRMap and the shadow store and flipping a control bit. Because the configuration is small (15\% per layer), it can be kept on-chip for both latency and integrity. The same mechanism applies to any fully connected layers in the MLP as well as to the Q/K/V projections in attention; in all cases, the accelerator sees a stream of matrix tiles and a stream of compact rewiring directives that gate only operand selection. The result is a security-aware inference flow that maintains near-baseline latency and energy while delivering the robustness benefits of FaR.

\begin{figure*}[!t]
\centering
\includegraphics[width=0.9 \textwidth] {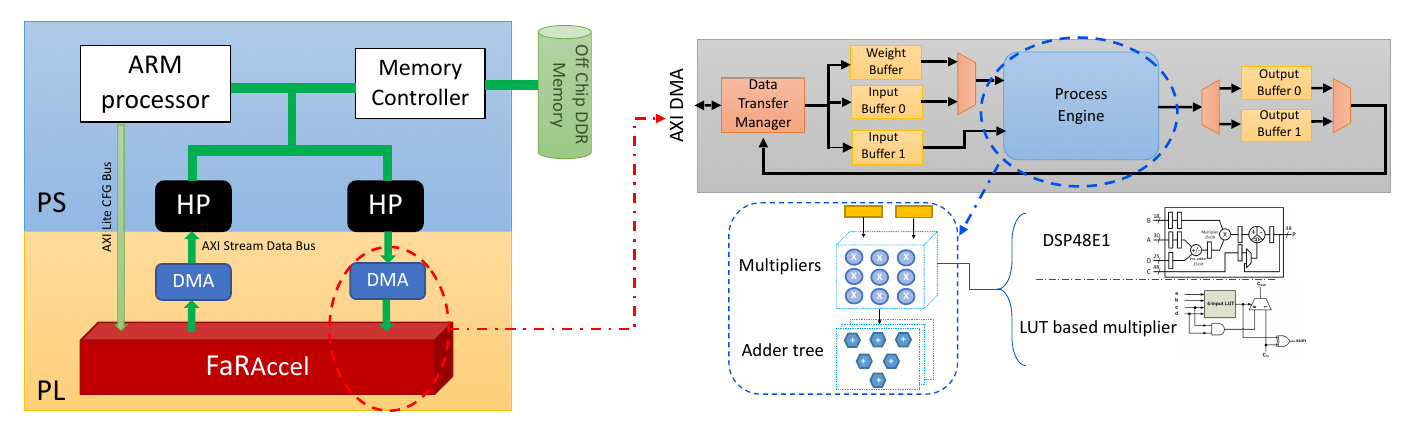}
\caption{Overview of the proposed accelerator on the FPGA device }
\label{fig:over}
\end{figure*}

\section{FaRAccel Implementation}

\subsection{Overview} FaRAccel is an FPGA-oriented accelerator composed of a grid of dot-product engines (DPEs), tile buffers for activations and baseline weights, a pair of small configuration memories for FaR metadata and pre-scaled donors, an operand redirect network integrated in front of each multiplier lane, and a controller that overlaps configuration decode with compute. The design keeps the compute datapath identical to a conventional GEMM core \cite{yi2024opengemm} and introduces FaR-specific logic only in the operand selection and control planes. Figure \ref{fig:over} presents the overview of our implementation.

\subsection{Dot-Product Engine (DPE)}
Each DPE is a 32-lane FP16 unit that computes one 32-element dot product per cycle after pipeline fill. A lane holds an FP16 multiplier, and the 32 products are reduced through a five-level pipelined adder tree with 31 FP16 adders. Small input registers capture the current activation and weight elements, and an output register holds the reduced result before write-back. The DPE consumes the inner-dimension stream while partial sums remain stationary in the reducer, which minimizes intermediate writes and matches the objective of constant throughput. Because the core compute is unchanged from a standard GEMM implementation, the DPE maps naturally to DSP blocks on FPGAs and to FP16 multiplier/adder macros on ASICs.

\subsection{FaR-Specific Logic}
The FaR-specific machinery sits immediately upstream of the multipliers. Two narrow SRAMs feed this machinery: the FaRMap cache and the shadow store. The FaRMap cache holds, for the 32 outputs in the current tile, a sparse list of victim–donor relations, division selectors, and skip flags. With the rewiring budget capped at 15\%, each output row typically has only a handful of entries. The shadow store holds only those donor weights that appear in the FaRMap, pre-scaled by one-half or one-third as required. To ensure the compute fabric never bubbles, the baseline weight buffer and the shadow store are read through independent ports so that a redirect decision does not create a structural hazard.

When the controller schedules an output row, it fetches the row’s FaRMap entries and expands them into a dense, 32-element select vector. Each vector element indicates whether a lane should read its weight from the baseline buffer, from a particular shadow address, or use a constant zero; this vector is then latched and remains stable while the row’s k-dimension is streamed. The per-lane operand redirect network is a small three-way selector that applies this decision before the multiplier. Because division factors are compiled into the shadow weights, the redirect network never performs arithmetic—its work is purely in choosing the correct operand source. The controller pipelines this select synthesis so that, while the current row is being reduced in the adder tree, it can already prepare the vector for the next row. This overlap hides the map-decode latency and allows the DPE to issue a new dot every cycle regardless of whether the next row contains rewiring.

\subsection{Memory Hierarchy and Data Movement}
The memory hierarchy and data movement follow standard accelerator practice, with double-buffered DMA moving activations and baseline weight tiles from DRAM into on-chip SRAMs and writing results back after accumulation. FaR data is minuscule compared to activations and weights, so it imposes negligible bandwidth pressure. 
Because the configuration footprint scales with the number of tiles rather than with the global layer size, even very large layers exhibit bounded on-chip state per active tile.

\subsection{Multi-PE Accelerator Design}
As a prototype, we are utilizing the design proposed in \cite{shen2017maximizing}, which is a layer-wise accelerator design for deep neural networks. Instead of solely utilizing Processing Engines (PEs) with varying kernel sizes, ours leverages PEs with a fixed quantization and precision scheme. 

We adopt a multiple PE accelerator approach. This approach enhances resource utilization and, in turn, the model's performance, as depicted in Figure \ref{fig:over}. The Zynq platform's processing system (PS) can interact with the Programmable Logic (PL) unit through high-performance ports linked to AXI DMAs. This connection provides the accelerator with high-bandwidth access to off-chip memory for read and write operations.
Each PE in the accelerator employs a ping-pong buffer for input and output values, while a singular buffer is used for weights. This arrangement stems from the fact that during a kernel's operation on inputs, weight values remain constant while inputs change following each kernel operation. Therefore, the frequency of weight transfers to and from the accelerator is considerably less than that of the inputs.


\subsection{Safety and Deployment Considerations}
Finally, FaRAccel includes safety hooks that are important for deployment. The controller validates FaRMap and shadow blobs before enabling a layer, and can fall back to baseline behavior if illegal indexes are fed or if configuration SRAMs indicate an error. Because FaR does not change the trained topology, a model can be deployed once and then hardened or relaxed over time by shipping new FaR configurations without touching weights or retraining. Taken together, these properties make the proposed architecture a practical path to bring FaR’s robustness into latency- and energy-constrained environments without compromising the performance characteristics of the underlying matrix engine.

\section{Evaluation}

In this section, we present hardware evaluation and performance results.

\subsection{Hardware Footprint, Timing, and Scalability}
FaRAccel is designed to preserve the compute macros of a standard GEMM core; hence, from a resource perspective, the architectural changes are modest. FaRAccel does not add multipliers or adders:  32-lane DPE still uses 32 FP16 multipliers and a 31-adder reduction tree. As a result, DSP usage remains essentially unchanged compared to the baseline. 
The additional logic consists of small per-lane multiplexers, lightweight control, and two compact memories for the FaR map and donor values. The added storage is on the order of one to one-and-a-half kilobytes for configuration and shadow values. In synthesis, these additions raise LUT and flip-flop counts only by a small percentage, while BRAM use increases modestly to provision the configuration and shadow stores; in our prototype, the configuration footprint is dominated by the on-chip tile buffers rather than by FaR-specific state. Power impact is similarly small because the selectors and configuration SRAMs toggle at row boundaries, whereas the multiplier and adder activity is identical to the baseline. Table \ref{table:util} presents the breakdown of FaRAccel resource utilization on our platform.

Timing closure mirrors that of the baseline because the operand-redirect network is placed one stage ahead of the multiplier inputs and is fully pipelined; consequently, the maximum clock frequency remains comparable to the GEMM-only design and is limited by the adder-tree depth rather than by FaR logic. Routing pressure is mitigated by keeping the select logic adjacent to the DSP columns and by duplicating registers on high-fan-out control nets; no cross-column long routes are introduced by FaR, which preserves slack on the critical paths. 

\begin{table}[b]
\vspace{-1em}
\caption{Breakdown of accelerator's resource utilization}
\centering
\scalebox{1.0}{
\begin{tabular}{|c|cccc|}
\hline
\multirow{2}{*}{\textbf{Component}} & \multicolumn{4}{c|}{\textbf{Hardware resources}}                                                                         \\ \cline{2-5} 
                                    & \multicolumn{1}{c|}{\textbf{LUT}} & \multicolumn{1}{c|}{\textbf{FF}} & \multicolumn{1}{c|}{\textbf{DSP}} & \textbf{BRAM} \\ \hline
\textbf{FIFOs}                      & \multicolumn{1}{c|}{465}          & \multicolumn{1}{c|}{330}         & \multicolumn{1}{c|}{0}            & 48            \\ \hline
\textbf{AXI DMAs}                   & \multicolumn{1}{c|}{6256}         & \multicolumn{1}{c|}{11048}       & \multicolumn{1}{c|}{0}            & 12            \\ \hline
\textbf{Compute engine}                   & \multicolumn{1}{c|}{1566}         & \multicolumn{1}{c|}{840}         & \multicolumn{1}{c|}{216}          & 0             \\ \hline

\textbf{Register File}              & \multicolumn{1}{c|}{0}            & \multicolumn{1}{c|}{19200}       & \multicolumn{1}{c|}{0}            & 0             \\ \hline
\textbf{Controller}                 & \multicolumn{1}{c|}{5311}         & \multicolumn{1}{c|}{3598}        & \multicolumn{1}{c|}{0}            & 0             \\ \hline
\textbf{Total}                      & \multicolumn{1}{c|}{11598}        & \multicolumn{1}{c|}{33016}       & \multicolumn{1}{c|}{216}          & 60            \\ \hline
\textbf{Available}                  & \multicolumn{1}{c|}{53200}        & \multicolumn{1}{c|}{106400}      & \multicolumn{1}{c|}{220}          & 140           \\ \hline
\textbf{Utilization}                & \multicolumn{1}{c|}{21.8\%}      & \multicolumn{1}{c|}{31.03\%}     & \multicolumn{1}{c|}{98.18\%}      & 42.86\%       \\ \hline
\end{tabular}
}
\label{table:util}
\end{table}

The design can scale along two orthogonal axes:  Within a processing element, lanes can be widened (e.g., to 64) to retire multiple dots per cycle, and the operand redirect network scales linearly with the number of lanes. Across the fabric, additional DPEs can be instantiated to process multiple tiles in parallel; each DPE carries its own small FaRMap cache and shadow store, so control remains local and contention-free. The same mechanism applies to both fully connected layers and to the Q/K/V projections in attention. Switching between these layer types is a matter of changing the stream of tiles and binding the appropriate FaRMap and shadow blobs; the compute remains the same.

\begin{figure}[!t]
\centering
\includegraphics[width=0.5 \textwidth] {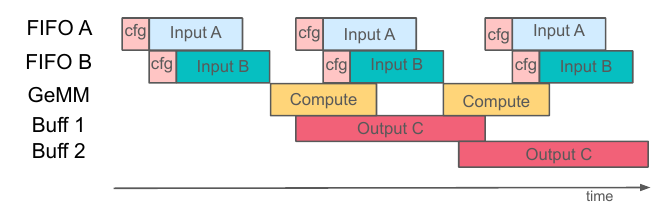}
\caption{Conceptual visualizations for configuration pre-loading, input pre-fetch, and output buffering }
\label{fig:pipe}
\end{figure}

From another perspective, spatial replication increases throughput linearly until it is bounded by off-chip memory bandwidth, while lane widening remains effective as long as on-chip SRAM banking can supply one FP16 operand per lane per cycle. In practice, double-buffered activation and weight tiles, combined with simple prefetching of the next FaR map block, keep the compute array fed for typical transformer shapes. Because FaR configuration is sparse and reused across many rows within a layer, the control plane consumes negligible bandwidth relative to activation and weight traffic and does not become a limiter at scale.
This is an important point since the steady-state throughput is preserved in our design.
This constant-throughput realization admits a simple performance model. With 32 lanes, a 32×32 tile produces 1,024 output dots and retires one dot per cycle after a 12-cycle pipeline fill, yielding 1,036 cycles per tile in the baseline. The FaR path adds one cycle per output row to latch the select vector, for an overhead of 32 cycles (about 3.1\%). The controller can overlaps select generation with the tail of the previous row’s reduction, so the common-case overhead is effectively zero. Figure \ref{fig:pipe} demonstrates the pipeline of FaRAccel.

\subsection{Performance Results}
We compare three realizations to isolate the effect of offloading FaR. The first baseline is a vanilla software stack without FaR that uses standard GEMM-backed linear layers in a deep learning framework. The second baseline implements FaR entirely in software and realizes rewiring by explicitly duplicating and scaling activations and by performing scatter/gather of weights, which prevents fusion into a single batched GEMM and introduces substantial framework overhead. Our method, FaRAccel, offloads only the FaR-aware linear projections while leaving all other layers and host orchestration identical to the software baselines; this ensures that any improvements stem from accelerating the FaR operations themselves rather than from unrelated changes in the model or runtime.

Three custom ViT structures trained MNIST, CIFAR-10, and CIFAR-100 datasets have been exploited for comparison. The details of custom models are presented in Table \ref{table:cus_vit}. 

By default, ViTs use normal $\texttt{linear()}$ functions, and their functionality is as simple as follows:

\begin{verbatim}
output = input.matmul(self.weight.t())
if self.bias is not None:
    output += self.bias
\end{verbatim}

While in a FaR-aware $\texttt{linear()}$ function, we should consider the extra FaR config and division factor that makes the software implementation very complex and inefficient, since we cannot use $\texttt{matmul}$ anymore for layers that have FaR config. However, on the hardware implementation, all the complexity of rewiring is handled by the hardware and hence, we can easily use the same DPE and $\texttt{matmul}$ function to perform the computation. The only thing that differs is the extra configuration loading introduced for multiplexing the shadow registers. 
For clarity, this work presents a FaR-aware accelerator \emph{prototype}: we offload only the FaR-enabled $\texttt{linear()}$ operator, not the entire model. Accordingly, our evaluation isolates and compares the performance overhead of the software FaR-aware $\texttt{linear()}$ against its hardware realization.

\begin{table}[b]
\caption{Custom ViTs for evaluation on selected Datasets}
\centering
\small
\vspace{1ex}
\scalebox{0.9}{
\begin{tabular}{|c|c|c|c|}
\hline
\textbf{ViT Params} & \textbf{MNIST} & \textbf{CIFAR-10} & \textbf{CIFAR-100} \\ \hline
image size          & 28             & 32                & 32                 \\ \hline
channel size        & 1              & 3                 & 3                  \\ \hline
patch size          & 7              & 8                 & 8                  \\ \hline
embed size          & 512            & 512               & 512                \\ \hline
num heads           & 8              & 8                 & 8                  \\ \hline
classes             & 10             & 10                & 100                \\ \hline
num layers          & 1              & 3                 & 6                  \\ \hline
hidden size         & 256            & 256               & 256                \\ \hline
Model size             & 3.19M              & 9.57M                 &  19.07M                 \\ \hline
\end{tabular}
}
\label{table:cus_vit}
\end{table}


Table \ref{table:micro} presents the results of our performance analysis with two FaR configurations. 
Microbenchmarks of the FaR-aware $\texttt{linear()}$ operator clarify where the speedup originates. We are observing 61\% overhead in $\texttt{matmul}$ operation when division factor is set to 3 while in FaRAccel the $\texttt{matmul}$ operation overhead is only 3\%. In software, the operator’s runtime scales approximately with the number of rewired inputs because the framework must materialize duplicated activations, perform per-lane scaling, and gather donor weights; these steps inhibit GEMM fusion and reduce library efficiency. In contrast, FaRAccel synthesizes the select vectors for each output row while the previous tile is reducing, and then presents operands directly to the multipliers without inserting arithmetic on the critical path. As a result, steady-state multiplier utilization matches that of a baseline GEMM on the same tile size, and the measured kernel latency approaches the baseline bound for memory-fed matrix multiply on our platform.

Across all three datasets, FaRAccel substantially reduces the overhead introduced by FaR when realized in software. When FaR is applied to a large fraction of linear layers, we observe the overhead drops from 15\% to 1\%, which shows up to a 10–15× end-to-end improvement relative to the FaR-software baseline, with the exact improvement depending on the fraction of rewired inputs per layer and the division factor used during activation splitting. These gains arise because FaRAccel converts rewiring into per-lane operand selection that runs at line rate and avoids fragmented GEMMs control flow. Accuracy under nominal (non-attack) evaluation remains matched to the FaR-software baseline, since the accelerator implements the same arithmetic at FP16 and the same sparse FaR map; in our experiments, the offload introduces no measurable deviation in top-1 accuracy.

We conduct two ablations to test sensitivity to FaR parameters and coverage. First, increasing the per-layer rewiring budget up to the default cap yields a near-linear runtime increase in software but leaves FaRAccel’s runtime comparatively flat, since per-lane selects are constant-time regardless of how many inputs are rewired. Second, moving from a division factor of two to three increases activation handling overhead in software yet has a negligible effect on FaRAccel, because the required scaling is compiled into the shadow donor values and does not execute at runtime. 

\begin{table}[t]
\caption{Performance analysis of FaRAccel with 15\% rewiring}
\centering
\small
\vspace{1ex}
\scalebox{0.9}{
\begin{tabular}{c|c|cc|cc|}
\cline{2-6}
                                              & \textbf{No FaR} & \multicolumn{2}{c|}{\textbf{SW FaR}} & \multicolumn{2}{c|}{\textbf{HW FaR}} \\ \cline{2-6} 
                                              & Baseline        & \multicolumn{1}{c|}{Div 2}  & Div 3  & \multicolumn{1}{c|}{Div 2}  & Div 3  \\ \hline
\multicolumn{1}{|c|}{\textbf{matmul latency}} & 1               & \multicolumn{1}{c|}{1.42}   & 1.61   & \multicolumn{1}{c|}{1.03}   & 1.03   \\ \hline
\multicolumn{1}{|c|}{\textbf{MNIST}}          & 1               & \multicolumn{1}{c|}{1.06}   & 1.07   & \multicolumn{1}{c|}{1.00}   & 1.00   \\ \hline
\multicolumn{1}{|c|}{\textbf{CIFAR 10}}       & 1               & \multicolumn{1}{c|}{1.09}   & 1.11   & \multicolumn{1}{c|}{1.01}   & 1.01   \\ \hline
\multicolumn{1}{|c|}{\textbf{CIFAR 100}}      & 1               & \multicolumn{1}{c|}{1.12}   & 1.15   & \multicolumn{1}{c|}{1.01}   & 1.01   \\ \hline
\end{tabular}
}
\label{table:micro}
\end{table}

\subsection{Limitations and Future Direction}
The current accelerator focuses on linear projections inside transformer blocks and does not accelerate convolutions or non-linearities; however, these components are not bottlenecks in the evaluated ViT models. Our evaluation targets a single FPGA device class and a fixed tile geometry; exploring alternative tilings, mixed-precision arithmetic, and multi-FPGA partitioning is left for future work. Finally, while the design supports replication of tiles and wider lanes, achieving near-linear scaling requires sufficient off-chip bandwidth and careful SRAM banking, which may need platform-specific tuning on devices with different memory organizations.

FaRAccel removes the software tax of FaR by realizing rewiring as constant-throughput operand selection in hardware. End-to-end inference time improves by as much as 10–15× over a FaR-in-software implementation while preserving nominal accuracy, and the hardware additions are modest enough to maintain the clock frequency and utilization characteristics of a baseline GEMM core.

\section{Related Works}

In this section, we introduce existing approaches of accelerating modern machine learning models and discuss their key features.

GPUs are commonly deployed as high-throughput accelerators in frameworks such as TensorFlow \cite{abadi2016tensorflow} and PyTorch \cite{paszke2019pytorch}. By exploiting massive parallelism and batching, they deliver strong throughput on a wide range of models. TensorRT \cite{zhou2022exploring} offers general-purpose mappings of deep learning graphs to GPUs, but it provides limited opportunities for workload-specific customization.
Gemmini \cite{genc2021gemmini} is an automatic accelerator generator capable of producing both systolic-array and parallel-vector architectures, and it has been widely used for deep learning workloads. For example, Sehoon et al. \cite{kim2023full} apply Gemmini to Transformer inference, characterizing Transformer behavior and introducing several optimizations. Other efforts tailor hardware specifically to Vision Transformers: ViTCoD \cite{you2023vitcod} designs a dedicated accelerator that handles both sparse and dense execution to raise utilization; Auto-ViT-Acc \cite{li2022auto} builds an FPGA accelerator for multi-head attention alongside an FPGA-aware quantization method; and HeatViT \cite{dong2023heatvit} targets embedded FPGAs with image-adaptive token pruning and 8-bit quantization. Despite these advances, most “sequential” accelerator flows still rely on a single, generic engine across layers with differing tensor shapes, which can lead to shape mismatches, under-utilization, and, ultimately, higher latency.

Unlike deep-learning training, real-time inference typically cannot aggregate inputs into large batches to expose extensive parallelism. Consequently, many batch-oriented, throughput-optimized systems utilize only a small fraction of their resources when serving a single request. Microsoft BrainWave \cite{fowers2018configurable} targets this production, datacenter-scale scenario by extracting intra-request (single-task) parallelism and mapping it efficiently onto FPGAs, achieving substantially lower latency than GPUs without sacrificing system-level throughput.
DNNExplorer \cite{zhang2020dnnexplorer} introduces a hybrid design methodology that deploys spatial accelerators for the initial layers and a generic accelerator for the remaining layers, enabling deeper networks while maintaining acceptable performance. However, DNNExplorer pipelines only between linear kernels, which reduces latency to a point;

Several works have extensively adapted machine-learning models to fit resource-constrained hardware \cite{ansarmohammadi2025mlb}, most notably through quantization \cite{nazari2019tot}, pruning, and related model-compression techniques \cite{nazari2020multi}. These approaches reduce on-device memory footprint and simplify arithmetic \cite{mirsalari2022fact}—e.g., replacing full-precision multiplications with shift/add or other bitwise operations—thereby improving efficiency \cite{nazari2023inter}. However, none of these efforts explicitly adopts a security-aware co-design to improve model or hardware resilience to fault mechanisms such as memory bit flips. To the best of our knowledge, our work is the first to introduce a high-level mitigation against bit-flip attacks that preserves performance while incurring only minimal hardware overhead.

\section{Conclusion}
\label{sec:conclusion}
In this paper, we introduced FaRAccel, the first hardware accelerator tailored to efficiently support the Forget and Rewire (FaR) defense against Bit-Flip Attacks (BFAs) in Transformer models. By reimagining FaR as an operand redirection problem rather than a graph rewrite, FaRAccel enables constant-throughput execution of rewired matrix operations with minimal hardware overhead. Our FPGA-oriented design integrates a lightweight configuration memory, a per-lane redirect network, and a shadow store for pre-scaled donor weights, all while preserving the original GEMM datapath and requiring no retraining or topology changes to the model. Evaluations show that FaRAccel maintains near-baseline latency and power efficiency and delivers the robustness benefits of FaR with less than 3\% worst-case overhead. This makes FaRAccel a practical and scalable solution for secure inference in edge and embedded AI deployments. Looking ahead, the principles established in this work open new opportunities for building resilience-aware hardware primitives that tightly couple algorithmic defenses with efficient architectural support. 

\bibliographystyle{IEEEtran}
\bibliography{bilbo}

\end{document}